\documentclass[10pt]{article}

\usepackage{a4wide,graphicx,pgf}
\usepackage{amsmath,amssymb,accents}
\usepackage{verbatim}

\title{\textbf{A note on spacelike and timelike compactness}}
\author{Ko Sanders}
\author{Ko Sanders\thanks{E-mail:
kosanders@uchicago.edu}\\
Enrico Fermi Institute\\
University of Chicago\\
5640 South Ellis Avenue\\
Chicago, IL 60637}
\date{11 November 2012}

\newtheorem{definition}{Definition}[section]
\newtheorem{theorem}[definition]{Theorem}
\newtheorem{proposition}[definition]{Proposition}
\newtheorem{corollary}[definition]{Corollary}
\newtheorem{lemma}[definition]{Lemma}
\newenvironment{proof*}{\smallskip\par\noindent\emph{Proof: }
 \ignorespaces}{\hfill$\Box$\smallskip\par\ignorespaces}

\newtheorem{example}[definition]{Example}

\newcommand{\map}[3]{\ensuremath{#1\!:\!#2\!\rightarrow\!#3}}

\begin{document}
\maketitle

\begin{abstract}
When studying the causal propagation of a field $\phi$ in a globally hyperbolic
spacetime $M$, one often wants to express the physical intuition that $\phi$ has
compact support in spacelike directions, or that its support is a spacelike
compact set. We compare a number of logically distinct formulations of this idea,
and of the complementary idea of timelike compactness, and we clarify their
interrelations. E.g., a closed set $A\subset M$ has a compact intersection with
all Cauchy surfaces if and only if $A\subset J(K)$ for some compact set $K$.
(However, it does not suffice to consider only those Cauchy surfaces that partake
in a given foliation of $M$.) Similarly, a closed set $A\subset M$ is contained
in a region of the form $J^+(\Sigma^-)\cap J^-(\Sigma^+)$ for two Cauchy surfaces
$\Sigma^{\pm}$ if and only if the intersection of $A$ with $J(K)$ is compact
for all compact $K$. We also treat future and past compact sets in a similar
way.
\end{abstract}

\section{Introduction}

Suppose $\phi$ is a physical field configuration on a globally hyperbolic
spacetime $M$, i.e.\ it is a (possibly distributional) section of some vector
bundle $V$ over $M$. When $\phi$ satisfies a normally hyperbolic equation of
motion with compactly supported initial data, then the support of $\phi$ is
contained in $J(K)$ for some compact $K\subset M$ and hence it has a compact
intersection with every Cauchy surfaces. Such solutions occur often in the
physics literature and are sometimes described as being ``compactly supported
on all Cauchy surfaces''. However, when $\phi$ is subject to a gauge symmetry,
the properties of $\phi$ are usually not uniquely determined by its initial
data, because one may always add gauge terms with largely uncontrolled
behaviour in the future or past. In this case it is less obvious whether the
criterion of compact support on all Cauchy surfaces still correctly encodes
the physical intuition that $\phi$ is ``spacelike compactly supported''. This
problem was encountered explicitly by \cite{Fewster+2012} in the context of
linearised general relativity. There the authors opted for the apparently
stronger criterion that $\phi$ has support in $J(K)$ for some compact
$K\subset M$.

In this note we will consider several distinct formulations of the idea that
$\phi$ has a spacelike compact support and we clarify their interrelations.
In particular we show the equivalence of the two formulations above (after
making them more precise). Furthermore, treating $\phi$ as a distribution
(density) and assuming it has a spacelike compact support, the natural class
of smooth testing sections of $V$ consists of the ones which have timelike
compact support. This leads us to consider also several distinct notions of
timelike compactness, in order to clarify their relations. In addition we
will take the time orientation of $M$ into account and treat future,
resp.\ past, compact supports along similar lines.

First, we consider a purely geometric situation, focussing on closed subsets
of $M$. In Sec.\ \ref{Sec_SC}, we discuss spacelike compact sets, together
with future and past spacelike compact sets. Sec.\ \ref{Sec_TC} deals with
timelike compact sets, together with future or past compact sets. After
these geometric preliminaries we consider in Sec.\ \ref{Sec_Supp} conditions
on distribution densities $\phi$ and on test-sections $f$, that guarantee
that their supports are spacelike compact. We also introduce natural
topologies on the spaces of future, past spacelike and timelike compactly
supported sections and distribution densities, so that they become each
others topological duals. We conclude our note in Section \ref{Sec_NHop} with
the special case where $\phi$ solves a normally hyperbolic equation and we
comment on the continuity of the unique advanced and retarded fundamental
solutions of such an operator w.r.t.\ the topologies on sections and
distributions with suitable supports.

Throughout we will use standard notions and notions from Lorentzian geometry
(e.g.\ \cite{ONeill}). Recall in particular that a Cauchy surface
$\Sigma\subset M$ is a subset which is intersected exactly once by every
inextendible timelike curve. We will assume that $M$ is globally hyperbolic,
which means that it has a Cauchy surface \cite{Bernal+2003}. In addition we
assume that a time-orientation for $M$ has been fixed. As a matter of
notation, we will let $\mathfrak{C}(M)$ denote the set of all Cauchy
surfaces in $M$ and $\mathfrak{C}_0(M)$ is the subset of all spacelike Cauchy
surfaces. The space of smooth sections of the vector bundle $V$ over $M$ will
be denoted by $\Gamma(M,V)$, while $\Gamma_0(M,V)$ denotes the space of
compactly supported smooth sections, both in their usual topologies
(cf.\ \cite{Baer+}). We let $\mathcal{D}(M,V^*)$ denote the space of
distribution densities with values in the dual vector bundle $V^*$ of $V$
(so that on an oriented spacetime $M$,
$\Gamma(M,V^*)\subset\mathcal{D}(M,V^*)$ by the natural pairing
$\langle\phi,f\rangle:=\int_M\phi(f)d\mathrm{vol}_g$, where $d\mathrm{vol}_g$
is the volume form induced by the metric $g$).

\section{Spacelike compact sets}\label{Sec_SC}
In this section we prove our main geometric result on spacelike compact sets
and its corollary on future and past spacelike compact sets. The technical
heart of these results is contained in the following proposition:
\begin{proposition}\label{Prop_SC}
Let $A\subset M$ be a closed set such that $A\cap\Sigma$ is compact (in
$\Sigma$, or, equivalently, in $M$) for all $\Sigma\in\mathfrak{C}_0(M)$.
Then there is a compact set $K\subset M$ such that $A\subset J(K)$.
\end{proposition}
A proof of this proposition is given at the end of this section. First,
however, we will discuss its consequences for spacelike compactness.

\begin{theorem}[Spacelike compact sets]\label{Thm_SC}
For a closed set $A\subset M$ in a globally hyperbolic spacetime the following
conditions are equivalent:
\begin{enumerate}
\item There is a compact set $K\subset M$ such that $A\subset J(K)$.
\item For every $\Sigma\in\mathfrak{C}(M)$, $A\cap\Sigma$ is compact.
\item For every $\Sigma\in\mathfrak{C}_0(M)$, $A\cap\Sigma$ is compact.
\end{enumerate}
\end{theorem}
Note in particular that this dispels the concern of \cite{Fewster+2012}
Footnote `b', that the first two items might not be equivalent.
\begin{proof*}
It is a well-known result in Lorentzian geometry that the first condition
implies the second (\cite{Baer+} Corollary A.5.4). The second implies the
third trivially and the third implies the first by Proposition \ref{Prop_SC}.
\end{proof*}

These results motivate the following definition:
\begin{definition}\label{Def_SCset}
We call a subset $A\subset M$ \emph{spacelike compact} when $\overline{A}$
satisfies any of the equivalent conditions of Theorem \ref{Thm_SC}.
\end{definition}

In Theorem \ref{Thm_SC} it does not suffice to consider only the Cauchy
surfaces of a given foliation of $M$. The following is an easy
counterexample:\footnote{\cite{Fewster2012}, Footnote 17, already gives a
counterexample consisting of a set $B\subset M$ which has compact
intersection with all Cauchy surfaces of a given foliation, but which is
not spacelike compact. However, that set $B$ is not closed and
$\overline{B}$ seems too pathological to occur as the support of a smooth
section.}
\begin{example}\label{Ex_notSC}
Consider the Minkowski spacetime $M_0$ in standard inertial coordinates
$(t,\mathbf{x})$ with $\mathbf{x}\in\mathbb{R}^{d-1}$ for some $d\ge 2$. We
use the foliation of $M_0$ by the constant $t$ Cauchy surfaces $\Sigma_t$.
For the set $A$ we choose the support of the function
$\phi(t,\mathbf{x}):=\psi(3e^{\|\mathbf{x}\|^2}t-3)$, where
$\psi\in C_0^{\infty}(\mathbb{R})$ has support $[-1,1]$. This means that
\[
A=\rm{supp}(\phi)=\left\{(t,\mathbf{x})|\ \frac23\le
e^{\|\mathbf{x}\|^2}t\le\frac43\right\}.
\]
It is easy to see that $A\cap\Sigma_t$ is compact for all $t\in\mathbb{R}$.
Now consider the hypersurface
$\Sigma:=\left\{(e^{-\|\mathbf{x}\|^2},\mathbf{x})\right\}$. One may
show that $\Sigma$ is a spacelike Cauchy surface (cf.\ \cite{Bernal+2003}
Corollary 11). To conclude the counterexample we note that
$\Sigma\subset A$, so $A\cap\Sigma=\Sigma$, which is not compact.
Hence, $A$ is not spacelike compact.
\hfill$\oslash$\smallskip\par\ignorespaces
\end{example}

Taking the time-orientation of $M$ into account we define the following
refined notions of spacelike compactness:
\begin{definition}\label{Def_fpsc}
We call a subset $A\subset M$ \emph{future}, resp.\ \emph{past},
\emph{spacelike compact} when $\overline{A}\subset J^-(K)$,
resp.\ $\overline{A}\subset J^+(K)$, for some compact $K\subset M$.
\end{definition}
Note that, informally speaking, the adjectives future, past and spacelike
refer to the regions of spacetime which do not intersect $A$. Future and
past spacelike compact sets are spacelike compact. A closed set is both
future and past spacelike compact if and only if it is compact.

\begin{corollary}\label{Cor_fpsc}
For a closed set $A\subset M$ the following conditions are equivalent:
\begin{enumerate}
\item $A$ is future (resp.\ past) spacelike compact.
\item $A\cap J^+(\Sigma)$ (resp.\ $A\cap J^-(\Sigma)$) is compact for every
$\Sigma\in\mathfrak{C}(M)$.
\item $A\cap J^+(\Sigma)$ (resp.\ $A\cap J^-(\Sigma)$) is compact for every
$\Sigma\in\mathfrak{C}_0(M)$.
\end{enumerate}
\end{corollary}
\begin{proof*}
It is well-known that the first condition implies the second (\cite{Baer+}
Corollary A.5.4). The second implies the third trivially. The third
condition implies that $A\cap\Sigma$ is compact for every
$\Sigma\in\mathfrak{C}_0(M)$, so $A\subset J(L)$ for some compact $L\subset M$,
by Proposition \ref{Prop_SC}. Furthermore, choosing a foliation of $M$ by
spacelike Cauchy surfaces $\Sigma_t$ (cf.\ \cite{Bernal+2006}) and using the
fact that for any $t\in\mathbb{R}$ the set $A\cap J^+(\Sigma_t)$
(resp.\ $A\cap J^-(\Sigma_t)$) is compact, we may find a $T$ such that
$A\subset J^-(\Sigma_T)$ (resp.\ $J^+(\Sigma_T)$). Choosing
$K:=J(L)\cap\Sigma_T$ we find $A\subset J^-(K)$ (res.\ $A\subset J^+(K)$),
proving the future (resp.\ past) spacelike compactness.
\end{proof*}

To conclude this section we supply the proof of Proposition \ref{Prop_SC}.
We begin with a lemma, which uses an exhaustion by compact sets
(\cite{Lee2011} Proposition 4.76):
\begin{lemma}\label{Lem_SC}
Let $\Sigma\in\mathfrak{C}_0(M)$ and let $\left\{K_n\right\}_{n\in\mathbb{N}}$
be an exhaustion of $\Sigma$ by compact sets, i.e.\ each $K_n\subset\Sigma$
is compact, $K_n\subset\accentset{\circ}{K}_{n+1}$ and
$\cup_{n\in\mathbb{N}}K_n=\Sigma$. Assume that there are sequences of
points $x_n\in M$ and compact spacelike acausal submanifolds $B_n\subset M$
with boundary, such that
\begin{enumerate}
\item $x_n\in B_n$,
\item $J(B_n)\cap\Sigma\subset\accentset{\circ}{K}_n$,
\item $J(B_{n+1})\cap K_n=\emptyset$.
\end{enumerate}
Then there is a $\Sigma'\in\mathfrak{C}_0(M)$ which contains all $B_n$, and
the set $X:=\bigcup_{n\in\mathbb{N}}\left\{x_n\right\}$ is closed, but not
compact.
\end{lemma}
\begin{proof*}
We may construct a spacelike Cauchy surface $\Sigma'\subset M$ that
contains all $B_n$ as follows. First we define $L_1:=K_1$ and by induction
we choose compact subsets $L_n\subset K_n$, $n\ge 2$, such that
$J(B_n)\cap\Sigma\subset\accentset{\circ}{L}_n$, but
$L_n\cap K_{n-1}=\emptyset$. (This is possible, by our assumptions on $B_n$
and $K_n$.) Note in particular that all $L_n$ are pairwise disjoint. The idea
is that the domain of dependence $M_n:=D(\accentset{\circ}{L}_n)$ provides
some room around $B_n$ to deform the Cauchy surface $\Sigma$, whilst the
$K_n$ ensure that the $B_n$ do not accumulate (see Figure \ref{Fig1}).

\begin{figure}
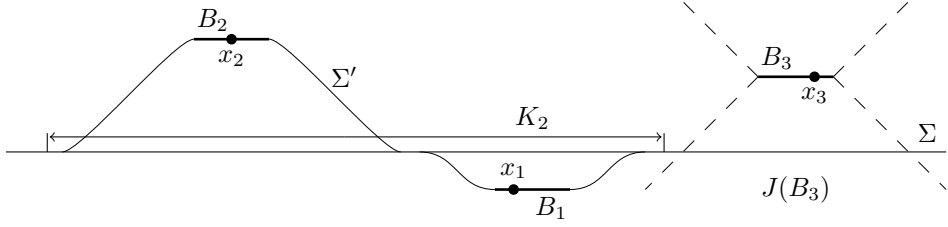
\label{Fig1}
\begin{center}
\begin{pgfpicture}{0cm}{1cm}{12.5cm}{4cm}
\color{black}
\pgfline{\pgfxy(0,2)}{\pgfxy(12.5,2)}
\pgfcircle[fill]{\pgfxy(3,3.5)}{2pt}
\pgfcircle[fill]{\pgfxy(6.75,1.5)}{2pt}
\pgfcircle[fill]{\pgfxy(10.75,3)}{2pt}
\pgfline{\pgfxy(0.55,2)}{\pgfxy(0.55,2.25)}
\pgfline{\pgfxy(8.75,2)}{\pgfxy(8.75,2.25)}
\pgfsetendarrow{\pgfarrowto}
\pgfline{\pgfxy(4.5,2.2)}{\pgfxy(8.7,2.2)}
\pgfline{\pgfxy(4.5,2.2)}{\pgfxy(0.6,2.2)}
\pgfclearendarrow
\pgfsetlinewidth{1pt}
\pgfline{\pgfxy(2.5,3.5)}{\pgfxy(3.5,3.5)}
\pgfline{\pgfxy(6.5,1.5)}{\pgfxy(7.5,1.5)}
\pgfline{\pgfxy(10,3)}{\pgfxy(11,3)}
\pgfsetlinewidth{0pt}
\pgfxycurve(0.75,2)(1,2)(2.25,3.5)(2.5,3.5)
\pgfxycurve(3.5,3.5)(3.75,3.5)(5,2)(5.25,2)
\pgfxycurve(5.5,2)(6,2)(6,1.5)(6.5,1.5)
\pgfxycurve(7.5,1.5)(8,1.5)(8,2)(8.5,2)
\pgfsetdash{{0.2cm}{0.2cm}}{0cm}
\pgfline{\pgfxy(10,3)}{\pgfxy(8.5,1.5)}
\pgfline{\pgfxy(11,3)}{\pgfxy(12.5,1.5)}
\pgfline{\pgfxy(10,3)}{\pgfxy(9,4)}
\pgfline{\pgfxy(11,3)}{\pgfxy(12,4)}
\pgfputat{\pgfxy(7.25,1.25)}{\pgfbox[center,center]{$B_1$}}
\pgfputat{\pgfxy(2.75,3.75)}{\pgfbox[center,center]{$B_2$}}
\pgfputat{\pgfxy(10.25,3.25)}{\pgfbox[center,center]{$B_3$}}
\pgfputat{\pgfxy(6.75,1.75)}{\pgfbox[center,center]{$x_1$}}
\pgfputat{\pgfxy(3,3.25)}{\pgfbox[center,center]{$x_2$}}
\pgfputat{\pgfxy(10.75,2.75)}{\pgfbox[center,center]{$x_3$}}
\pgfputat{\pgfxy(7,2.45)}{\pgfbox[center,center]{$K_2$}}
\pgfputat{\pgfxy(10.5,1.5)}{\pgfbox[center,center]{$J(B_3)$}}
\pgfputat{\pgfxy(12.25,2.25)}{\pgfbox[center,center]{$\Sigma$}}
\pgfputat{\pgfxy(4.5,3)}{\pgfbox[center,center]{$\Sigma'$}}
\end{pgfpicture}
\end{center}
\caption{A schematic depiction of the geometric construction
used to prove Lemma \ref{Lem_SC}.}
\end{figure}

For each $n\in\mathbb{N}$ the region $M_n$ is a globally hyperbolic spacetime
in its own right (\cite{Baer+} Lemma A.5.9). We
may choose spacelike Cauchy surfaces $S_n$ for $M_n$ such that $B_n\subset S_n$,
by \cite{Bernal+2006} Theorem 1.1. We then set
\[
\Sigma':=\left(\Sigma\setminus\bigcup_{n\in\mathbb{N}}
\accentset{\circ}{L}_n\right)\cup\bigcup_{n\in\mathbb{N}}S_n.
\]
To prove that $\Sigma'$ is a Cauchy surface for $M$ we let $\gamma$ be an
arbitrary inextendible timelike curve. If $\gamma$ does not intersect $\Sigma$
in some $\accentset{\circ}{L}_n$, then it already intersects $\Sigma'$.
Moreover, this point of intersection is unique, as $\gamma$ cannot intersect
any of the $M_n$. On the other hand, if $\gamma$ intersects some
$\accentset{\circ}{L}_n$, then it cannot intersect $\Sigma'$ in
$\Sigma\cap\Sigma'$ or in any $S_k$ with $k\not=n$. Furthermore, $\gamma$
intersects $M_n$ and the intersection is an inextendible causal curve in $M_n$,
which has a unique point of intersection with $S_n$. Therefore, $\Sigma'$ is a
Cauchy surface. Also note that $\Sigma'$ contains all $B_n$, by construction,
and that it is spacelike, because $\Sigma$ is spacelike.

To conclude the proof we show that $X\subset\Sigma'$ is closed but not
compact. First suppose that $y\in\overline{X}$ and let $U\subset\Sigma'$ be
a compact neighbourhood of $y$. Note that $J(U)\cap\Sigma\subset K_N$ for
some $N\in\mathbb{N}$. By construction, $K_N$ does not intersect $L_n$ with
$n>N$, so $D(K_N)\cap\Sigma'$ does not contain $x_n$ with $n>N$. It follows
that $y$ must be one of the points $x_1,\ldots,x_N$, so $X$ is closed. Now
consider the open cover of $X$ consisting of the sets
$\left\{S_n,n\in\mathbb{N}\right\}$. Each $x_n$ is contained only in the
corresponding $S_n$, so there is no proper subcover. This proves in
particular that there is no finite subcover, so $X\subset\Sigma'$ is not
compact.
\end{proof*}

We may now prove Proposition \ref{Prop_SC}:
\begin{proof*}
We will assume that there is no set $K$ such that $A\subset J(K)$ and derive
a contradiction. For this purpose we fix a $\Sigma\in\mathfrak{C}_0(M)$ and
an exhaustion of $\Sigma$ by compact sets
$\left\{K_j\right\}_{j\in\mathbb{N}}$. We consider the set
$\dot{A}:=A\setminus\Sigma$ and note that $\dot{A}$ is not contained in any
set of the form $J(L)$ with compact $L\subset\Sigma$ (otherwise we could
take $K=L\cup (A\cap\Sigma)$). In particular, $\dot{A}\not=\emptyset$, so
we may choose $x_1\in\dot{A}$ and $j_1\in\mathbb{N}$ such that
$x_1\in D(\accentset{\circ}{K}_{j_1})$. We now proceed by induction to
choose sequences of points $x_n\in\dot{A}$ and numbers $j_n\in\mathbb{N}$
such that $x_n\in D(\accentset{\circ}{K}_{j_n})$ and
$J(x_{n+1})\cap K_{j_n}=\emptyset$. This is possible, because for each $n$,
$\dot{A}\setminus J(K_{j_n})$ contains some point $x_{n+1}$ and the compact
set $J(x_{n+1})\cap\Sigma$ is contained in the interior of some
$K_{j_{n+1}}$.

Note that $n\mapsto j_n$ is strictly increasing, so $K_{j_n}$ is again an
exhaustion of $\Sigma$ by compact sets. Using this in Lemma \ref{Lem_SC}
with $B_n:=\left\{x_n\right\}$ yields a spacelike Cauchy surface $\Sigma'$
containing all $x_n$, but for which $A\cap\Sigma'\supset X$ is not compact.
This is the desired contradiction.
\end{proof*}

\section{Timelike compact sets}\label{Sec_TC}

We now turn to the complementary notion of timelike compact sets. In this
case our main geometric result is
\begin{theorem}[Future and past compact sets]\label{Thm_TC}
For a closed set $A\subset M$ in a globally hyperbolic spacetime the following
conditions are equivalent:
\begin{enumerate}
\item There is a Cauchy surface $\Sigma\subset M$ such that
$A\subset J^+(\Sigma)$ (resp.\ $A\subset J^-(\Sigma)$).
\item For every compact set $K\subset M$, the set $A\cap J^-(K)$
(resp.\ $A\cap J^+(K)$) is compact.
\item For every point $p\in M$, the set $A\cap J^-(p)$
(resp.\ $A\cap J^+(p)$) is compact.
\end{enumerate}
\end{theorem}
\begin{proof*}
For any compact set $K\subset M$ and any Cauchy surface
$\Sigma\subset M$, the sets $J^{\pm}(K)$ are closed and the intersection
$J^{\pm}(K)\cap J^{\mp}(\Sigma)$ is compact (cf.\ \cite{Baer+} Lemma A.5.4,
Lemma A.5.1 and the comment above Lemma A.5.7). It then follows immediately
that the first condition implies the second. The second implies the third
trivially. It only remains to show that the third condition implies the
first.

By a reversal of time-orientation it suffices to consider the case where
$A\cap J^-(p)$ is compact for all $p\in M$. We choose a global time
function $t$ on $M$ and a foliation $M\simeq\mathbb{R}\times\Sigma$ by
Cauchy surfaces, so that $t$ is the projection onto the first factor
(cf.\ \cite{Bernal+2006}). For each inextendible timelike curve $\gamma$
in $M$ we then define
\[
t_-(\gamma):=\min\left\{0; t(x), x\in\gamma\cap J^+(A)\right\}
\]
The minimum $t_-(\gamma)$ exists, because if $x\in\gamma\cap J^+(A)$,
then $\gamma\cap J^+(A)\cap J^-(x)$ is compact and $t_-(\gamma)$ is the
minimum value of $t$ on this set.

Now consider the inextendible timelike curves $\gamma_p(t):=(t,p)$,
define $T_-(p):=t_-(\gamma_p)$ and consider the embedding
$\map{\psi_-}{\Sigma}{M}$ by $\psi_-(p):=(T_-(p),p)$. The image $\Sigma_-$
of $\psi_-$ has the following properties. Firstly, if $(t,p)\in A$, then
$T_-(p)\le t$ by construction, so $A\subset J^+(\Sigma_-)$. Secondly,
$\Sigma_-$ is achronal, for if there were a timelike curve $\gamma_1$
between, say, $(T_-(p),p)$ and $(T_-(q),q)$ with $T_-(q)\ge T_-(p)$, and
if $\gamma_2$ is a causal curve from some point $x\in A$ to $(T_-(p),p)$,
then the concatenation of $\gamma_1$ and $\gamma_2$ can be deformed to a
time-like curve from $x$ to $(T_-(q),q)$ (cf.\ \cite{ONeill}). Hence,
$T_-(q)$ cannot be the minimum as defined, leading to a contradiction.
(If no such $\gamma_2$ exists, then $T_-(p)=T_-(q)=0$ and $\gamma_1$
cannot exist either.) Thus we see that $\Sigma_-$ is achronal. Finally,
$\Sigma_-$ is a Cauchy surface. To prove this we consider an inextendible
causal curve $\tau\mapsto\gamma(\tau)$ in $M$. There is a unique point
$p\in\Sigma$ such that $(t_-(\gamma),p)\in\gamma$. Both when
$\gamma\cap J^+(A)=\emptyset$ and when $\gamma\cap J^+(A)\not=\emptyset$
one may see that $(t_-(\gamma),p)\in\Sigma_-$, by an argument that
involves the concatenation of causal curves as above, together with the
definition of $T_-(p)$. Therefore, $\gamma$ intersects $\Sigma_-$, and
as $\Sigma_-$ is achronal, the point of intersection is unique. This
proves that $\Sigma_-$ is a Cauchy surface with $A\subset J^+(\Sigma_-)$,
so we established the first condition.
\end{proof*}

\begin{definition}\label{Def_TCset}
We call a subset $A\subset M$ \emph{future}, resp.\ \emph{past}, \emph{compact}
when there is a Cauchy surface $\Sigma\subset M$ such that $A\subset J^-(\Sigma)$,
resp.\ $A\subset J^+(\Sigma)$. We call $A$ \emph{timelike compact} when $A$
is both future and a past compact.
\end{definition}
By Theorem \ref{Thm_TC}, our definition of future and past compact sets
is equivalent to the one in \cite{Baer+}, at least for closed subsets
of globally hyperbolic spacetimes.
Using the same theorem it may easily be shown that a set is future,
resp.\ past, spacelike compact if and only if it is both spacelike compact
and future, resp.\ past, compact (cf.\ the proof of Corollary
\ref{Cor_fpsc}).

When $A\subset M$ is timelike compact and we consider a foliation of $M$ by
Cauchy surfaces $\Sigma_t$, it is not necessarily true that there are
numbers $t_-<t_+$ such that $A\subset J^+(\Sigma_{t_-})\cap J^-(\Sigma_{t_+})$.
A counterexample in Minkowski spacetime can be obtained, using the notations
of Example \ref{Ex_notSC}, by choosing $A$ to be the image of $\Sigma_0$
under a non-trivial Lorentz boost. Clearly $A$ itself is still a Cauchy
surface and hence timelike compact, but it contains points with arbitrary
values of $t$.

Note furthermore that in order to establish timelike compactness it does
not suffice that $A$ has a compact intersection with all inextendible
causal curves. The following is a counterexample:
\begin{example}\label{Ex_notTC}
Consider the Minkowski spacetime $M_0$ in standard inertial coordinates
$(t,\mathbf{x})$ with $\mathbf{x}\in\mathbb{R}^{d-1}$ for some $d\ge 2$.
The region $M':=I^+(0)\subset M$ is a globally hyperbolic spacetime in its
own right and the hypersurfaces
$\Sigma_R:=\left\{t=\sqrt{R^2+\|\mathbf{x}}\|^2\right\}$, $R>0$, foliate
$M'$ by Cauchy surfaces.
Note that $M'$ cannot contain a Cauchy surface for $M$, because for any
$\mathbf{x}$ with unit norm, the inextendible timelike curve
$\gamma_{\mathbf{x}}(\tau):=(\sinh(\tau),\cosh(\tau)\mathbf{x})$ does not
enter $M'$. For the set $A$ we choose the support of the function
$\phi(t,\mathbf{x}):=\psi(2\sqrt{1+\|\mathbf{x}\|^2}(t-\sqrt{1+\|\mathbf{x}\|^2}))$,
where $\psi\in C_0^{\infty}(\mathbb{R})$ has support $[-1,1]$. Note that
$A$ is timelike compact in $M'$ (using the foliation). However, it cannot
be timelike compact in $M$, because the inextendible timelike curve
$\gamma_{\mathbf{x}}$ lies entirely in $J^-(A)\setminus A$. Hence, if
$A\subset J^-(\Sigma)$ for some Cauchy surface $\Sigma$ and if
$x\in\gamma_{\mathbf{x}}\cap\Sigma$, we could construct a timelike curve
from $x$ via $A$ to $\Sigma$, contradicting the fact that $\Sigma$ is
Cauchy. Nevertheless, any inextendible causal curve $\gamma$ has a compact
intersection with $A$, because if $\gamma$ does not enter $M'$ the
intersection is empty, while if $\gamma$ does enter $M'$, the intersection
is compact, since $A$ is timelike compact in $M'$.
\hfill$\oslash$\smallskip\par\ignorespaces
\end{example}

\section{Spacelike and timelike compact supports}\label{Sec_Supp}

Now we return to the original motivation and consider a distribution
density $\phi$ with values in some vector bundle $V$ on $M$. We make
the following obvious definition:
\begin{definition}
A distribution density $\phi$ on $M$ is said to have spacelike, timelike,
future (spacelike), resp.\ past (spacelike) compact support if and only
if $\mathrm{supp}(\phi)$ is spacelike, timelike, future (spacelike),
resp.\ past (spacelike) compact.
\end{definition}
Again, it does not suffice to consider only a particular foliation of
Cauchy surfaces to obtain spacelike compactness, nor does it suffice to
assume compact intersections with all inextendible causal curves to obtain
timelike compactness. Indeed, both of the counterexamples \ref{Ex_notSC}
and \ref{Ex_notTC} are based on the supports of smooth sections $\phi$.
However, in the spacelike case we do have the following result:
\begin{theorem}\label{Thm_SCsupp}
Let $\phi$ be a distribution density on $M$ and assume that either
\begin{enumerate}
\item[a)] $\phi$ is continuous, or
\item[b)] $WF(\phi)$ has no timelike vectors, so its restriction to all
spacelike Cauchy surfaces is well-defined by microlocal arguments.
\end{enumerate}
Then the following conditions are equivalent:
\begin{enumerate}
\item $\phi$ is spacelike compactly supported.
\item There is a compact set $K\subset M$ such that
$\mathrm{supp}(\phi|_{\Sigma})\subset J(K)$ for all
$\Sigma\in\mathfrak{C}(M)$.
\item There is a compact set $K\subset M$ such that
$\mathrm{supp}(\phi|_{\Sigma})\subset J(K)$ for all
$\Sigma\in\mathfrak{C}_0(M)$.
\item $\mathrm{supp}(\phi|_{\Sigma})$ is compact for all
$\Sigma\in\mathfrak{C}(M)$.
\item $\mathrm{supp}(\phi|_{\Sigma})$ is compact for all
$\Sigma\in\mathfrak{C}_0(M)$.
\end{enumerate}
\end{theorem}
\begin{proof*}
The implications 2$\rightarrow$3 and 4$\rightarrow$5 are trivial. The
implications 2$\rightarrow$4 and 3$\rightarrow$5 follow from the fact that
$J(K)\cap\Sigma$ is compact for every compact $K\subset M$ and every Cauchy
surface $\Sigma\subset M$ (\cite{Baer+} Lemma A.5.4). Furthermore,
1$\rightarrow$2 follows from Theorem \ref{Thm_SC} and the fact that
$\mathrm{supp}(\phi|_{\Sigma})\subset\mathrm{supp}(\phi)\cap\Sigma$. To
complete the proof it suffices to prove that 5$\rightarrow$1. By Theorem
\ref{Thm_SC} we only need to show that $\mathrm{supp}(\phi)\cap\Sigma$ is
compact for all $\Sigma\in\mathfrak{C}_0(M)$. We will argue by contradiction,
so we assume that there is a spacelike Cauchy surface $\Sigma\subset M$
such that $\mathrm{supp}(\phi)\cap\Sigma$ is not compact. We may foliate
$M$ by spacelike Cauchy surfaces $\Sigma_t$, $t\in\mathbb{R}$, such that
the projection $t$ on the first factor is a global time coordinate and
$\Sigma=\Sigma_0$ (cf.\ \cite{Bernal+2006} Theorem 1.2).

We can find an exhaustion of $\Sigma$ by compact sets $K_n$ and a
sequence of points $x_n\in\Sigma$ such that $x_n\subset\accentset{\circ}{K}_n$
and $x_{n+1}\not\in K_n$, much in the same way is in the proof of
Proposition \ref{Prop_SC}. We now write $x_n=(0,q_n)$ and recall that
$x_n\in\mathrm{supp}(\phi)$. For any open neighbourhood $U\subset\Sigma$
of $q_n$ and any $\epsilon>0$ we may choose a test-function
$\chi\in C_0^{\infty}(U)$ such that the distribution density
$t\mapsto\phi(t,\chi)$ does not vanish identically on
$(-\epsilon,\epsilon)$, by Schwartz' Kernels Theorem. Furthermore, by
assumption a) or b) this distribution is at least continuous, so there
is some $t_n\in(-\epsilon,\epsilon)$ for which $\phi(t_n,\chi)\not=0$.
This entails that $(t_n,q_n)\in\mathrm{supp}(\phi|_{\Sigma_{t_n}})$.

By induction we choose a sequence of numbers $\epsilon_n>0$ which is
sufficiently small to ensure that
$J(\pm\epsilon_n,q_n)\subset\accentset{\circ}{K}_n$ and
$J(\pm\epsilon_{n+1},q_{n+1})\cap K_n=\emptyset$ for all $n$ and both
signs. Then, choosing $t_n\in(-\epsilon_n,\epsilon_n))$ as above, we
may choose compact subsets $B_n\subset\Sigma_{t_n}$ such that
$J(B_n)\cap\Sigma\subset\accentset{\circ}{K}_n$ and
$J(B_{n+1})\cap K_n=\emptyset$. With these $x_n$, $B_n$ and $K_n$ the
assumptions of Lemma \ref{Lem_SC} are satisfied, so there is a
spacelike Cauchy surface $\Sigma'$ containing all $B_n$ and such that
the set $X:=\cup_{n\in\mathbb{N}}\left\{x_n\right\}$ is closed but not
compact in $\Sigma'$. Since $\Sigma'$ and $\Sigma_{t_n}$ coincide in a
neighbourhood of $x_n$, $x_n$ is also in
$\mathrm{supp}(\phi|_{\Sigma'})$. In other words,
$\mathrm{supp}(\phi|_{\Sigma'})\supset X$ and therefore
$\mathrm{supp}(\phi|_{\Sigma'})$ is not a compact set. This
contradicts the assumptions, hence $\phi$ must have spacelike compact
support.
\end{proof*}

For any closed set $B\subset M$ we may consider the space $\Gamma(B,V)$ of
smooth sections of $V$ on $M$ with support in $B$, as a closed subspace of
$\Gamma(M,V)$. In analogy to $\Gamma_0(M,V)$ we may then define the spaces
of sections with spacelike, timelike and future, resp.\ past, (spacelike)
compact supports as inductive limits (cf.\ \cite{Schaefer}):
\begin{eqnarray}
\Gamma_{fsc}(M,V):=\bigcup_{K\subset M}\Gamma(J^-(K),V),&&
\Gamma_{fc}(M,V):=\bigcup_{\Sigma\subset M}\Gamma(J^-(\Sigma),V),\nonumber\\
\Gamma_{psc}(M,V):=\bigcup_{K\subset M}\Gamma(J^+(K),V),&&
\Gamma_{pc}(M,V):=\bigcup_{\Sigma\subset M}\Gamma(J^+(\Sigma),V),\nonumber\\
\Gamma_{sc}(M,V):=\bigcup_{K\subset M}\Gamma(J(K),V),&&
\Gamma_{tc}(M,V):=\bigcup_{\Sigma^{\pm}\subset M}
\Gamma(J^+(\Sigma^-)\cap J^-(\Sigma^+),V),\nonumber
\end{eqnarray}
where $K$ is compact and $\Sigma,\Sigma^{\pm}$ are Cauchy surfaces. (For
the spacelike compact case this agrees with Definition 3.4.6 of
\cite{Baer+}. For smooth functions the topologies on $C^{\infty}_{sc}(M)$,
$C^{\infty}_{fsc}(M)$ and $C^{\infty}_{psc}(M)$ coincide with those
introduced by \cite{Ferguson2012}.) With these topologies, the following
inclusions are continuous
\begin{eqnarray}\label{Eq_CtsIncl}
&&\Gamma_0(M,V)\subset\Gamma_{fsc}(M,V)\subset\Gamma_{sc}(M,V)
\subset\Gamma(M,V),\\
&&\Gamma_0(M,V)\subset\Gamma_{tc}(M,V)\subset\Gamma_{fc}(M,V)
\subset\Gamma(M,V),\nonumber
\end{eqnarray}
and similarly with past (spacelike) compact instead of future (spacelike)
compact supports.

In an analogous way we may introduce spaces of distribution densities with
the same support properties, which will be indicated by the same subscripts,
e.g.
\[
\mathcal{D}_{sc}(M,V)=\bigcup_{K\subset M}\mathcal{D}(J(K),V),
\]
where $\mathcal{D}(B,V)$ is the space of distribution densities with support
in $B$, as a closed linear subspace of $\mathcal{D}(M,V)$ in the usual
distributional topology.

\begin{theorem}
Each of the spaces $\Gamma_*(M,V)$, where $*$ indicates any of the subscripts
$fsc$, $psc$, $sc$, $fc$, $pc$, $tc$, is reflexive and we have
\begin{eqnarray}
\mathcal{D}_{fsc}(M,V^*)=\Gamma_{pc}(M,V)'&
\mathcal{D}_{psc}(M,V^*)=\Gamma_{fc}(M,V)'&
\mathcal{D}_{sc}(M,V^*)=\Gamma_{tc}(M,V)'\nonumber\\
\mathcal{D}_{fc}(M,V^*)=\Gamma_{psc}(M,V)'&
\mathcal{D}_{pc}(M,V^*)=\Gamma_{fsc}(M,V)'&
\mathcal{D}_{tc}(M,V^*)=\Gamma_{sc}(M,V)'.\nonumber
\end{eqnarray}
\end{theorem}
\begin{proof*}
Using the continuous embeddings in equation (\ref{Eq_CtsIncl}), any
$\phi\in\Gamma_*(M,V)'$ is a distribution density. In the case $*=pc$, let
$\Sigma\subset M$ be any Cauchy surface. The restriction map from
$\Gamma(J^+(\Sigma),V)$ to $\Gamma(I^+(\Sigma),V)$ is continuous and it
has a dense range, as may be shown by direct approximation, using
multiplication with suitable cut-off functions. Therefore, the restriction
of $\phi$ to $I^+(\Sigma)$ is continuous on $\Gamma(I^+(\Sigma),V)$, so it
has compact support. It follows that $I^+(\Sigma)\cap\mathrm{supp}(\phi)$
is compact for any $\Sigma$ and hence
$I^+(\Sigma)\cap\mathrm{supp}(\phi)$ is compact too (since
$J^+(\Sigma)\subset I^+(\Sigma')$ for some $\Sigma'$). By Corollary
\ref{Cor_fpsc} $\phi$ has future spacelike compact support. Conversely, if
$\phi$ has future spacelike compact support, then we can find a smooth
cut-off function $\chi\in C^{\infty}_{fsc}(M)$ such that $\chi\equiv 1$ on
$\mathrm{supp}(\phi)$. The map $f\mapsto \chi f$ is continuous from
$\Gamma_{pc}(M,V)$ to $\Gamma_0(M,V)$ and $\phi(f)=\phi(\chi f)$, so
$\phi\in\Gamma_{pc}(M,V)'$.

The second item on the first line is proved by reversing the
time-orientation. The third item is proved in a similar way, using Theorem
\ref{Thm_SC} instead of Corollary \ref{Cor_fpsc}. The items on the second
line are also proved in a similar way, but now using Theorem \ref{Thm_TC}.
%
%
Finally we note that both $\Gamma(M,V)$ and $\Gamma_0(M,V)$ are reflexive.
The reflexivity of all $\Gamma_*(M,V)$ then follows from the proofs above,
if we interchange the roles of smooth sections and distribution densities.
\end{proof*}

\section{Consequences for normally hyperbolic operators}\label{Sec_NHop}

To conclude this note we consider the case where $\phi$ satisfies a
linear, normally hyperbolic field equation. In this case one expects
that the spacelike compactness is preserved under the time evolution,
so it would suffice to consider only one Cauchy surface. To be more
precise,
\begin{proposition}
If $\phi$ satisfies a normally hyperbolic equation, then the following
are equivalent:
\begin{enumerate}
\item $\phi$ has spacelike compact support.
\item $\mathrm{supp}(\phi|_{\Sigma})$ is compact for all
$\Sigma\in\mathfrak{C}(M)$.
\item There is a smooth spacelike Cauchy surface $\Sigma\in\mathfrak{C}_0(M)$
such that $\mathrm{supp}(\phi)\cap\Sigma$ is compact.
\end{enumerate}
\end{proposition}
\begin{proof*}
We have already seen in Theorem \ref{Thm_SCsupp} that the first and second
items are equivalent and they both trivially imply the third. For the
converse one uses the well-posedness of the Cauchy problem and the fact
that compactness of $\mathrm{supp}(\phi)\cap\Sigma$ implies that both
initial data on $\Sigma$ have compact support.
\end{proof*}
Note that in this case it does suffice to consider the Cauchy surfaces
$\Sigma_t$ which belong to a given foliation of $M$ and to require that
$\mathrm{supp}(\phi)\cap\Sigma_t$ is compact. It clearly does not suffice
to require that $\phi|_{\Sigma}$ has compact support for a single spacelike
$\Sigma\in\mathfrak{C}(M)$, because the other initial datum may not have
compact support. However, it is less clear whether it suffices to require
that $\mathrm{supp}(\phi|_{\Sigma_t})$ is compact for all $t\in\mathbb{R}$
and a given foliation $\Sigma_t$ of $M$.
%

Let $P$ denote a normally hyperbolic operator in the vector bundle $V$ over
$M$ and let $E^{\pm}$ denote the unique advanced and retarded fundamental
operators. It is well-known \cite{Baer+} that these are continuous linear
maps
\[
\map{E^{\pm}}{\Gamma_0(M,V)}{\Gamma_{sc}(M,V)}
\]
such that $\mathrm{supp}(E^{\pm}f)\subset J^{\pm}(\mathrm{supp}(f))$. Using
the topologies introduced in Section \ref{Sec_Supp} and the support
properties it is in fact not hard to show that the maps
\begin{eqnarray}
\map{E^+}{\Gamma_{psc}(M,V)}{\Gamma_{psc}(M,V)}&&
\map{E^-}{\Gamma_{fsc}(M,V)}{\Gamma_{fsc}(M,V)}\nonumber\\
\map{E^+}{\Gamma_{pc}(M,V)}{\Gamma_{pc}(M,V)}&&
\map{E^-}{\Gamma_{fc}(M,V)}{\Gamma_{fc}(M,V)}\nonumber
\end{eqnarray}
are continuous. (The proof is analogous to that of \cite{Ferguson2012} Lemma
3.11). This entails e.g.\ that $\map{E^+}{\Gamma_0(M,V)}{\Gamma_{psc}(M,V)}$
and $\map{E^+}{\Gamma_{tc}(M,V)}{\Gamma_{pc}(M,V)}$ are also continuous,
by the continuous inclusions (\ref{Eq_CtsIncl}). When $M$ is oriented one may
define the operators $E^+$ also on distributional sections, by duality. We
then have $(E^{\pm}\phi,f)=(\phi,E^{\mp}f)$, which leads to continuous linear
maps
\begin{eqnarray}
\map{E^+}{\mathcal{D}_{psc}(M,V)}{\mathcal{D}_{psc}(M,V)}&&
\map{E^-}{\mathcal{D}_{fsc}(M,V)}{\mathcal{D}_{fsc}(M,V)}\nonumber\\
\map{E^+}{\mathcal{D}_{pc}(M,V)}{\mathcal{D}_{pc}(M,V)}&&
\map{E^-}{\mathcal{D}_{fc}(M,V)}{\mathcal{D}_{fc}(M,V)}.\nonumber
\end{eqnarray}

\section*{Acknowledgements}
I am grateful to Chris Fewster for useful comments and suggestions.


\begin{thebibliography}{15}
\bibitem{Baer+}
C.\ B\"ar, N.\ Ginoux and F.\ Pf\"affle, \emph{Wave equations on Lorentzian manifolds and quantization},
ESI Lectures in mathematics and physics, EMS, Z\"urich 2007

\bibitem{Bernal+2003}
A.\ N.\ Bernal and M. S\'anchez, Commun.\ Math.\ Phys.\ \textbf{243}, 461--470 (2003)

\bibitem{Bernal+2006}
A.\ N.\ Bernal and M. S\'anchez, Lett.\ Math.\ Phys.\ \textbf{77}, 183--197 (2006)

\bibitem{Ferguson2012}
M. Ferguson, arXiv:1203.2151v1 [math-ph]

\bibitem{Fewster+2012}
C.\ J.\ Fewster and D.S. Hunt, arXiv:1203.0261v3 [math-ph]

\bibitem{Fewster2012}
C.\ J.\ Fewster, \emph{Lectures on quantum energy inequalities},
arXiv:1208.5399v1 [gr-qc]

\bibitem{Lee2011}
J.\ M.\ Lee, \emph{Introduction to topological manifolds},
Springer, New York 2011

\bibitem{ONeill}
B.\ O'Neill, \emph{Semi-Riemannian geometry}, Academic Press, New York 1983

\bibitem{Schaefer}
H.\ H.\ Schaefer, \emph{Topological vector spaces}, Springer, New York 1999
\end{thebibliography}
\end{document}